\documentclass[prl,showpacs]{revtex4}
\usepackage{graphicx}
\usepackage{amsfonts}

\begin{document}


\title{Stochastic incoherence in the response of rebound bursters}

\author{Marzena Ciszak}

\affiliation{CNR-Istituto Nazionale di Ottica, Largo E. Fermi 6,
50125 Firenze, Italy}

\begin{abstract}
At an optimal value of the noise intensity, the maximum
variability in rebound burst durations is observed and referred to
as a response stochastic incoherence. A general mechanism
underlying this phenomenon is given, being different from those
reported so far in excitable systems. It is shown to be determined
by (i) the monotonous reduction of the hysteresis responsible for
bursting caused by noise and consequent transformation of
responses from rebound bursts to single spikes, and (ii) a
symmetry breaking in distributions of burst durations caused by
existence of the minimum response length. The phenomenon is
studied numerically in a Morris-Lecar model for neurons and its
mechanism is explained with the use of canonical models describing
hard excitation states.
\end{abstract}



\maketitle


\section{Introduction}
The presence of noise in nonlinear systems has been shown to
induce non-trivial behavior in their dynamics~\cite{lefever}.
Stochastic resonance~\cite{jung} occurs when the system is driven
by both a periodic signal and noisy fluctuations. At an optimal
value of the noise amplitude, the system exhibits the maximum
correlation with the periodic signal. This phenomenon has been
exhaustively studied theoretically and demonstrated experimentally
in different kinds of nonlinear systems~\cite{30}, e.g. in
excitable~\cite{franci} and bistable~\cite{bistSR} ones. It may
occur, however, that even in the absence of periodic forcing the
system reveals coherent oscillations at an optimum noise
amplitude. This is the case of stochastic coherence (or coherence
resonance)~\cite{haken,4,orderNoise}. The reason for occurrence of
stochastic coherence in excitable systems has been attributed to
the existence of two different characteristic
times~\cite{pakdaman}, activation and excursion times, which
induce a coherence maximum in the inter spike intervals when they
both reach a mutual minimum. Another proposed mechanism given
in~\cite{stone} associates the appearance of coherence maximum
with the noise induced bifurcation from the excitable to
oscillatory regime. Stochastic coherence has been demonstrated
theoretically in many systems including excitable~\cite{3} and
autonomous bursting systems~\cite{hindRose,cohResML}. It has also
been demonstrated experimentally in various physical
systems~\cite{23,25,29}.

While stochastic resonance can appear in systems exhibiting
various dynamical regimes, stochastic coherence can appear only in
excitable systems, since it requires the existence of a refractory
period. Excitability is a crucial feature of neurons which enables
an easy and efficient interaction between them~\cite{izhikReview}.
In two-dimensional excitable systems, an above-threshold
perturbation applied at time $t=t_0$, triggers a large excursion
in phase space, until it finally returns back to the stable fixed
point. During the spike duration the system is unsensitive to
external perturbations. This temporal period marks a refractory
period of the system. Higher-dimensional systems can give rise to
a different type of excitability, namely, excitable or rebound
bursting. The system being in the state of excitable bursting (at
$t>t_0$), with the difference to standard excitable systems (e. g.
FitzHugh-Nagumo model~\cite{fhn}), is susceptible to noisy
fluctuations. Thus it has rather a pseudo-refractory period
(pseudo, since it can be modified by the presence of noise). On
the other side, in the absence of noisy fluctuations at $t>t_0$,
the excitable burst has a well defined width determined by system
parameters.

\begin{figure}[h]
\begin{center}
\includegraphics*[width=0.8\columnwidth]{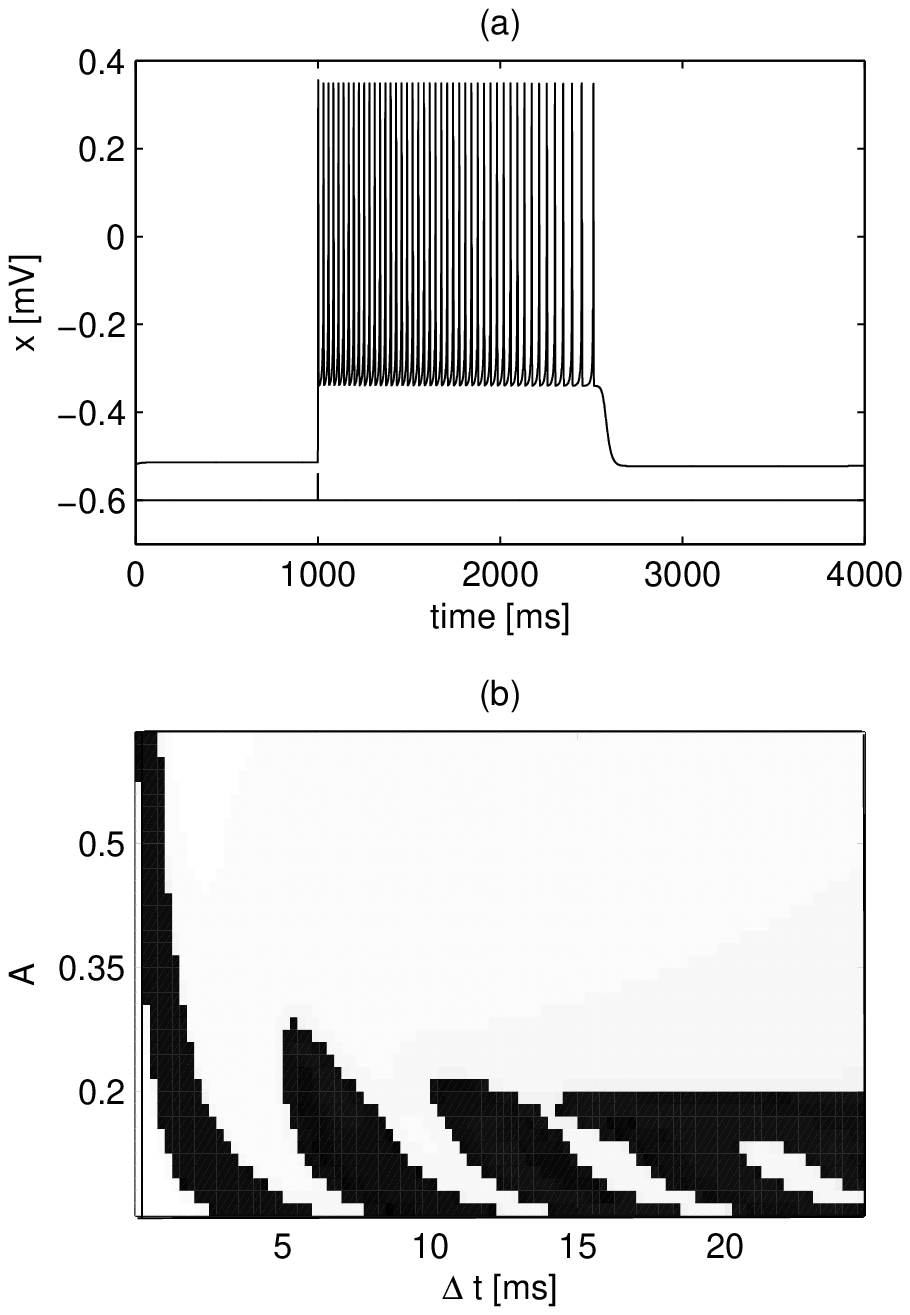}
\caption{\label{fig1} (a) Excitable burst (upper time series)
evoked by noiseless external stimulation $S$ (lower time series
plotted out of scale). (b) Response characteristics of the system
to external stimulations of amplitude $A$ and duration $\Delta t$.
White color corresponds to a single spike response, and black
color corresponds to fixed duration bursts. In (a) and (b) the
Morris-Lecar system with $c=0.28$ is considered.}
\end{center}
\end{figure}

In addition to the stochastic coherence, the phenomenon of
stochastic incoherence (SI) has been reported, showing that in
excitable systems exists a range of noise amplitudes at which the
maximum inter-spike variability is observed. The phenomenon has
been demonstrated in leaky integrate-and-fire model~\cite{lindner}
where the mechanism of occurrence has been attributed to the third
time scale existing in excitable systems, the absolute refractory
period. SI has been also shown to occur in FitzHugh-Nagumo
model~\cite{sancho} where the underlying mechanism has been
associated with the nonmonotonous relaxational behaviour of the
system near the oscillatory regime.

In this work a different type of stochastic incoherence is
reported, namely, {\it response } stochastic incoherence (RSI).
Its occurrence is found in a dynamical regime of excitable
bursting. In the case of two dimensional excitable systems the
spike duration (refractory period) is approximately fixed while
the inter-spike interval is extremely sensitive to noise. In the
case of bursting systems subject to noise, maximal variability is
observed both in the inter-burst intervals and in burst durations.
However, the two phenomena are observed in different range of
noise amplitudes. In particular, the maximal variability in the
inter-burst interval occurs for higher noise levels and refers to
stochastic incoherence. In this study, the generation of bursts is
triggered by a stimulus and not by the accompanying noise, whose
only role is that of inducing variability in the duration of the
elicited burst. It is shown, that the variability of excitable
burst durations $T_B$ manifests a pronounced maximum at the
optimal amplitude of fluctuations. The phenomenon is referred to
as an {\it incoherent} in order to maintain a close relation with
a well-known SI observed in excitable systems. The meaning of an
incoherence in the case of RSI is associated with a maximum
irregularity observed in $T_B$ as the noise intensity is varied.
It is demonstrated that the mechanism of the variability
maximization in response duration is mainly determined by the
sensitivity of the system hysteresis, which determines bursting
regimes, to noise. During a decrease of the mean burst duration,
or equivalently, a decrease of the hysteresis depth, the system
transforms from hard excitation state (hysteretic) into a soft
excitation state (non-hysteretic), leading to a single spike mean
responses at larger noise amplitudes. The maximum variability
occurs when the symmetry breaking in the distributions for burst
durations takes place. This symmetry breaking is caused by the
existence of the minimum possible response length (single spike)
which becomes highly probable at larger noise amplitudes. The
occurrence of the phenomenon is demonstrated in the Morris-Lecar
model equations. Its underlying mechanism is explained on examples
of canonical systems containing hysteresis. In particular, in
models describing stabilized subcritical pitchfork bifurcation (or
subcritical Hopf bifurcation), and in a model describing a damped
pendulum.

\section{Model}
The general model equations are considered:
\begin{eqnarray}
\{\epsilon _1 \dot{x},\epsilon _2 \dot{y},\epsilon
_3\dot{z}\}=\{f(x,y,z)+ S(t), g(x,y), h(x,z)\} \label{eq1}
\end{eqnarray}
where $x-y$ is the fast subsystem and $z$ is the slow variable.
The fast subsystem $x-y$ gives autonomous spiking, meanwhile
variable $z$ rebounds spikes and enables bursting regimes,
including excitable bursts. $S(t)$ is an external perturbation
acting as an external stimulus provoking the excitable response of
a neuron and has the form of the Dirac $\delta $ function. One
chosen system parameter is then assumed to fluctuate following an
independent white noise process $\xi(t)$ with correlations
$\langle \xi(t)\xi(t')\rangle=2D\delta(t-t')$ and zero mean. The
choice of the random process type is motivated by results reported
in Ref.~\cite{rowat} which show that Gaussian noise has the same
effect as channel or synaptic noise.

\begin{figure}[h]
\begin{center}
\includegraphics*[width=0.75\columnwidth]{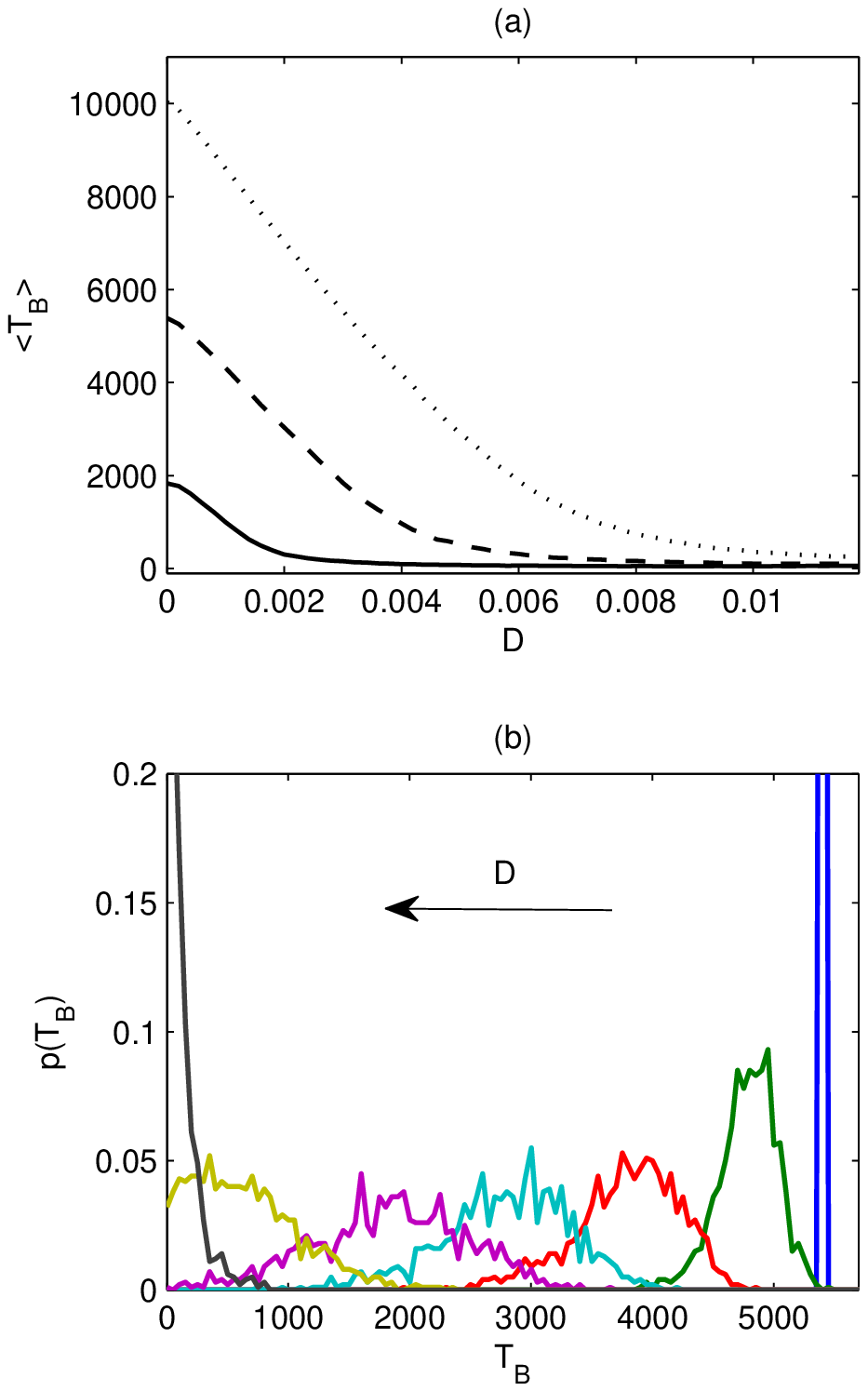}
\caption{\label{fig2} (a) Mean response durations for different
values of the control parameter: $c=0.291$ (solid line), $c=0.326$
(dashed line) and $c=0.368$ (dotted line). (b) Normalized
distributions of the response durations in the case of $c=0.326$
shown for increasing noise amplitudes (from right to left):
$D=\{0,6\times 10^{-4}, 0.0014, 0.0022, 0.003, 0.0046, 0.0118\}$
(the distributions for $D=0$ and $D=0.0118$ are shown partially).}
\end{center}
\end{figure}

\subsection{Rebound bursting in Morris-Lecar system}

A special case of the general model in Eq.~(\ref{eq1}) is the
modified Morris-Lecar model~\cite{lecar}:
\begin{eqnarray}
\epsilon _1 \dot{x}&=&\Gamma
(x)-2y(x+0.7)-0.01z(x+0.885)+S(t)\nonumber
\\
\epsilon _2 \dot{y}&=&c(-y+1/(1+e^{-(x-0.1)/0.07})\nonumber \\
\epsilon _3\dot{z}&=&-z+9.3(x+0.6)\label{eq2}
\end{eqnarray}
where $\Gamma
(x)=-(1+e^{(-(x+0.01)/0.075)})^{-1}(x-1)-0.065(x+0.5)$,
$\epsilon_1=0.97$, $\epsilon_2=1/\cosh((x-0.1)/0.14)$ and
$\epsilon_3=30550$. The Morris-Lecar model is a conductance-based
model of voltage oscillations in the barnacle giant muscle fiber
which is often used as a qualitatively accurate model of neuronal
spiking. In order to obtain a rebound burst, the additional
variable $z$ is added to the subsystem $x-y$. An example of such a
bursting response is shown in Fig.~\ref{fig1} (a). The system in
Eq.~(\ref{eq2}) contains a dynamical hysteresis, i.e. a regime
with transient bistability between a fixed point and repetitive
spiking. The existence of such a dynamical hysteresis is
responsible for bursting dynamics. The rest state disappears via
fold bifurcation, and the periodic spiking disappears via
saddle-homoclinic orbit bifurcation~\cite{terman}. The latter
bifurcation allows class I neural excitability, since it manifests
in low frequencies appearing at the transition to a rest state
onset.

As already mentioned in the Introduction, the rebound bursting
system has a rather pseudo-refractory period, since it can be
influenced by the slight external fluctuations. On the other hand,
the external stimulation has to be strong enough to elicit the
rebound bursting response. Let consider the noiseless case when
the system is stimulated with a rectangular single pulse of
amplitude $A$ and duration $\Delta t$. Numerical simulations show
that the system responses have a form of a fix duration burst or
single spike. The nonlinear responses characteristics to such
external stimulation are shown in Fig.~\ref{fig1} (b). It can be
seen that the response of the system is bistable (coexistence of
single spike and fixed duration bursting responses). A lack of
variability in the burst durations and the two state responses of
the system to stimuli with different parameters ($A$ and $\Delta
t$) suggest, that the system is rather unsensitive to fluctuations
in external stimulation eliciting excitable burst. As a
consequence, only fluctuations present in an upper spiking state
can give rise to a variability in the response durations. In the
presence of noiseless stimulus $S$, the system fires a burst of a
fixed duration, with a well-defined passage from the lower to the
upper state and {\it vice versa}, determined by internal system
parameters. In Eq.~(\ref{eq2}) the burst duration can be
controlled by modulation of the parameter $c$.

\subsection{RSI in rebound burster}

\begin{figure}[h]
\begin{center}
\includegraphics*[width=1.\columnwidth]{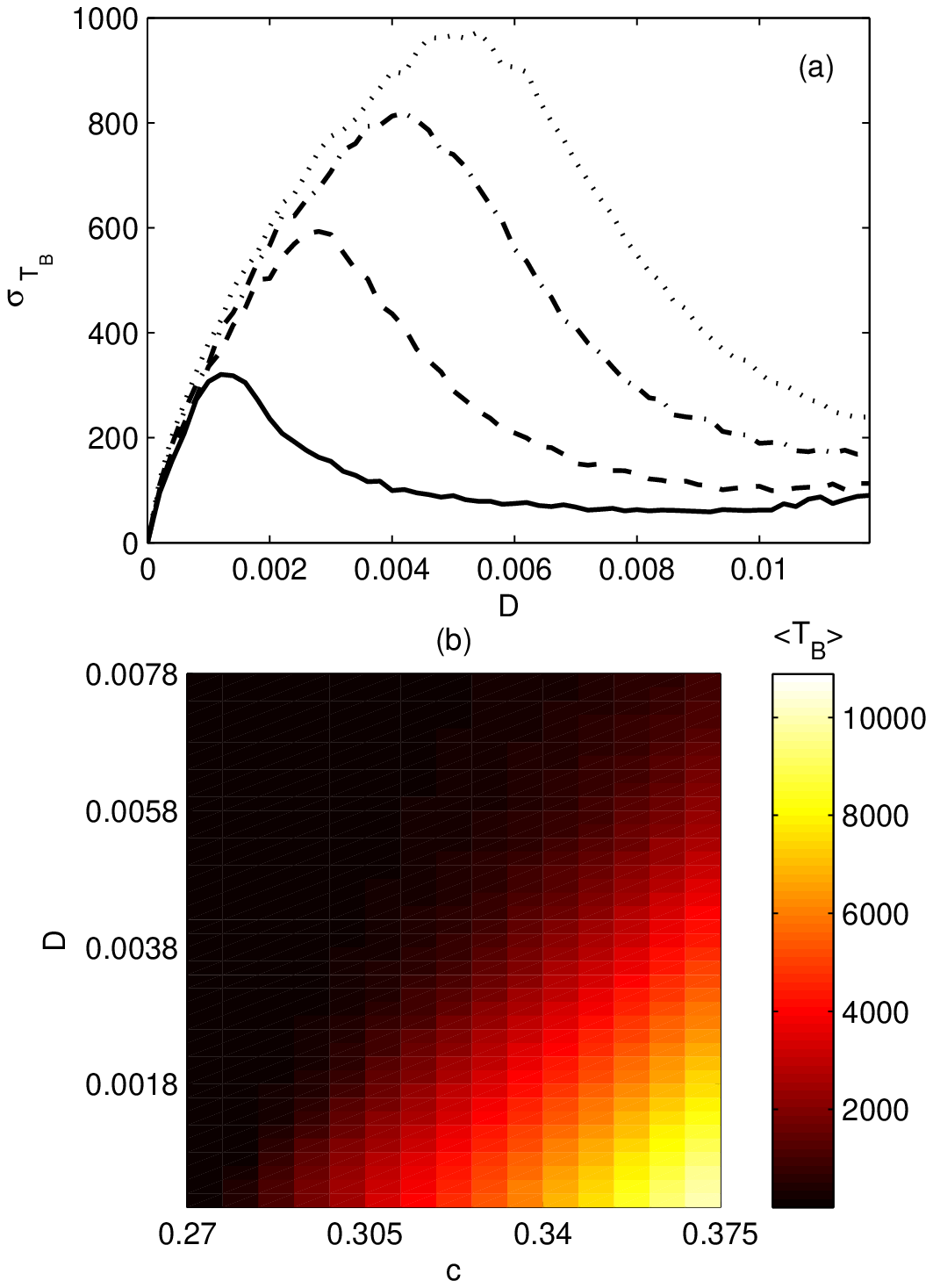}
\caption{\label{fig3} (a) Standard deviation $\sigma_ {T_B}$
versus noise amplitude $D$ for different values of the control
parameter: $c=0.291$, $c=0.319$, $c=0.347$ and $c=0.368$. (b)
Dependence of $\langle T_B\rangle $ on noise amplitude $D$ and
control parameter $c$.}
\end{center}
\end{figure}

Two cases are considered: i) the parameter $c$ is chosen to
fluctuate $c\rightarrow c+D\xi(t)$, modelling the internal noise
or ii) the stimulus $S$ is chosen to be accompanied by external
fluctuations $S\rightarrow S+D\xi(t)$. The rebound bursting can be
measured in different manners. One way is to estimate the total
time during which the repetitive spiking takes place. Another way
is to count the number of spikes in a given burst, as usually done
in experiments~\cite{experRebound}. In the present work, the
former way for measuring the burst durations $T_B$ is used,
namely, the estimation of temporal intervals during which the
system fires spikes repetitively. In order to describe
qualitatively the variability of the response durations, standard
deviation is considered $\sigma _{T_B}=(\langle
T_B^2\rangle-\langle T_B\rangle ^2)^{\frac{1}{2}}$ where $\langle
\rangle$ stands for mean. The statistical measurements are done on
an ensemble of excited bursts in the presence of different
realizations of noise. The noise intensities taken into
consideration are small enough to keep $FAR=0$ (FAR-false alarm
rate) at $S=0$. This is done in order to study the specific
response of the system to external stimulation. It is observed,
that the mean burst duration decreases with increasing of noise
intensity. In Fig.~\ref{fig2} (a) the effect of noise on the mean
burst duration for different values of the parameter $c$ is shown.
This implies that in the presence of noise, the passage from the
upper to the lower state becomes faster, and the transition time
decreases as noise amplitude increases. The same behaviour is
observed for both types of fluctuations present in the system. In
view of the above results, it can be hypothesized that noise
destroys the transiently existing bistability region in the
bursting system, and transforms the hard excitation system into a
standard excitable one-spike system or soft excitation system.
This is clearly shown in Fig.~\ref{fig2} (b) where the
distributions of the burst durations are plotted for different
values of noise amplitude. From these distributions $\sigma
_{T_B}$ is calculated and plotted in Fig.~\ref{fig3} (a) for
various noise intensities $D$. The maximum of $\sigma _{T_B}$
corresponds to a maximum variation in excitable burst durations.
The choice of $\sigma _{T_B}$ to describe the variability of $T_B$
instead of the usually used coefficient of variation $CV=\sigma
_{T_B}/<T_B>$ is caused by the fact that $<T_B>$ decreases with
the increase of $D$. Thus, having a division by decreasing
quantity, $CV$ hides the useful information about $T_B$
variability. In fact, it has been already pointed out
in~\cite{lindner2002}, that $CV$ sometimes fails as an indicator
of coherence.

It has been shown that the burst duration can be modified either
by modulation of a control parameter $c$, either by varying the
noise intensity. In fact, both, the effects of a control parameter
$c$ and noise amplitude $D$ on the burst duration, are shown to be
equivalent (see Fig.~\ref{fig3} (b)). This implies that random
fluctuations in the rebound burster can play a role of a control
parameter.

\section{Mechanism of RSI}
\subsection{Sensitivity of hard excitation states to noise}

\begin{figure}[h]
\begin{center}
\includegraphics*[width=1.\columnwidth]{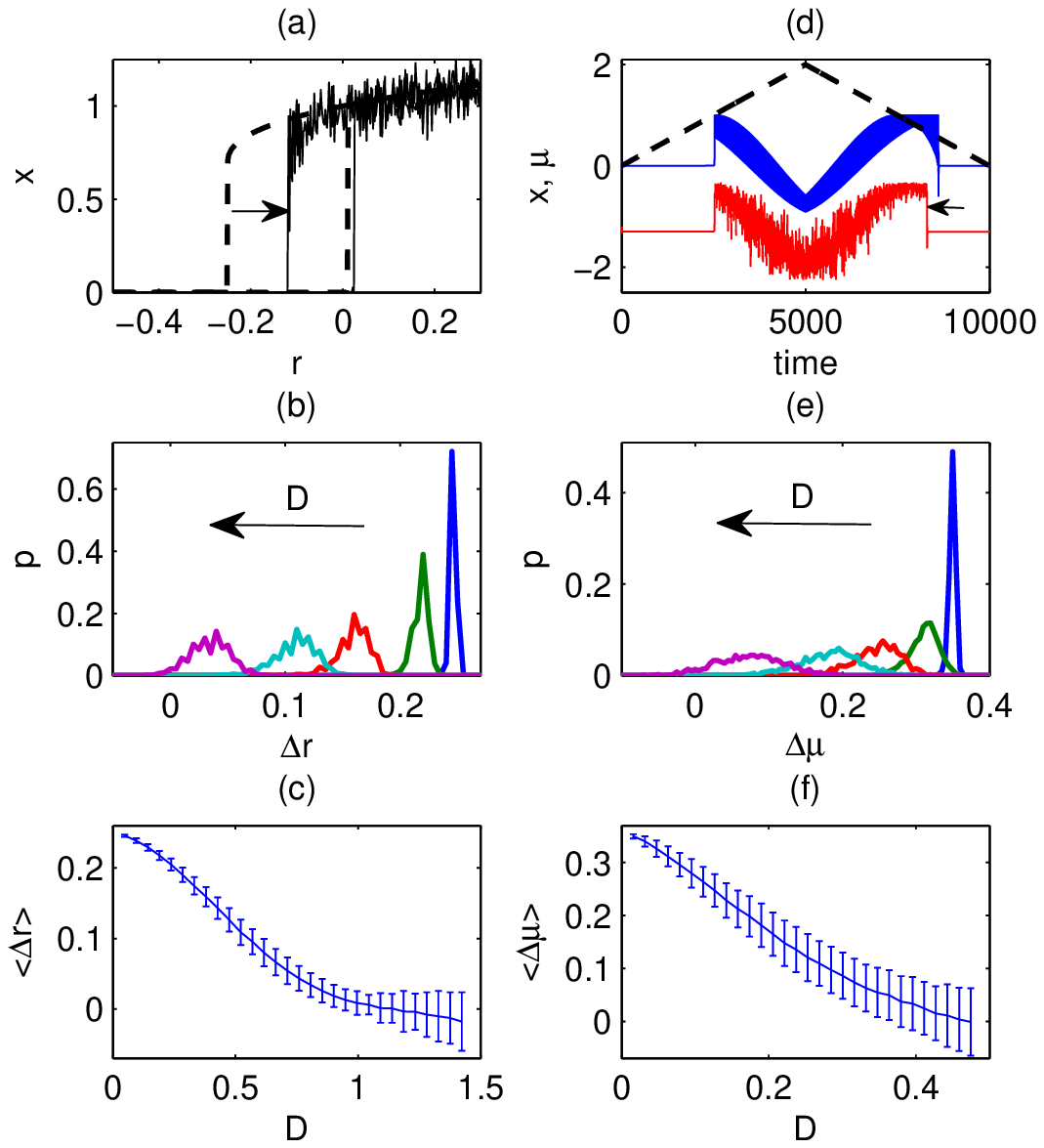}
\caption{\label{fig4} A canonical model for stabilized subcritical
pitchfork bifurcation: (a) bifurcation diagram for a system
without (dashed line) and with noise (solid line); the arrow marks
the hysteresis reduction; (b) normalized distributions of the
hysteresis lengths $\Delta r$ for selected noise amplitudes (from
right to left): $D=\{0.0474, 0.1897, 0.3795, 0.5218, 0.8064\}$;
(c) mean hysteresis length $\langle \Delta r\rangle $ and its
standard deviation $\sigma _{\langle \Delta r\rangle}$. Model
describing damped pendulum: (d) response of the system to a
variation of internal parameter $\mu $ (dashed line); upper trace
corresponds to the system without noise and the lower to the
system with noise; the arrow shows the hysteresis reduction; (e)
normalized distributions of the hysteresis lengths $\Delta \mu$
for selected noise amplitudes (from right to left): $D=\{0.0158,
0.0632, 0.1265, 0.1897, 0.3162\}$; (f) analogue to (c).}
\end{center}
\end{figure}

In order to understand the effects of noise on the burst duration
in the Morris-Lecar system let consider a canonical model of a
system containing hysteresis which describes a stabilized
subcritical pitchfork bifurcation~\cite{strogatz}:
\begin{eqnarray}
\dot{x}=rx+x^3-x^5 \label{pitch}
\end{eqnarray}
In the case, when the additional subsystem equation $\dot{\phi
}=2\pi$ is added to Eq.~(\ref{pitch}) and $x$ is considered as a
radial variable, the system undergoes subcritical Hopf
bifurcation~\cite{mackey}. In the absence of the stabilizing term
$x^5$ the system in Eq.~(\ref{pitch}) undergoes a supercritical
pitchfork bifurcation when one unstable fixed point bifurcates to
three fixed points, one stable and two unstable. The stabilizing
term allows the re-occurrence of the two stable and one unstable
fixed points. The common feature between the saddle-node
(occurring in the Morris-Lecar model) and stabilized subcritical
pitchfork bifurcations is the existence of low frequencies (in the
case of saddle-node bifurcation) or low amplitudes (in the case of
pitchfork bifurcation) appearing at the bifurcation onset. The two
backward-bending branches of unstable fixed points bifurcate from
the origin when $r=r_0=0$. Due to a stabilizing term $x^5$, the
unstable branches bend and become stable at $r=r_s$, where
$r_s<0$. The stable large-amplitude branches exist for all
$r>r_s$. In the range $r_s<r<0$, two different stable states
coexist, the origin and the large-amplitude fixed points, which
marks the hysteresis region. Inside this hysteretic region, the
initial condition $x_0$ determines which fixed point is approached
as $t\rightarrow \infty $. In the white noise environment, the
following stochastic differential equation is obtained for
Eq.~(\ref{pitch}):
\begin{eqnarray}
dx_t=(r x_t+x_t^3-x_t^5)dt+D x_t dW_t\label{hysterEq}
\end{eqnarray}
where the parameter $r$ is assumed to fluctuate as $r_t=r+D\xi_t$,
where $r$ is an average value, $\xi_t$ is a Gaussian white noise
and $D$ is the intensity of the fluctuations. Numerical
simulations of Eq.~(\ref{hysterEq}) for different realization of
noises reveal, that the same phenomenon as in the Morris-Lecar
system occurs: the reduction of the mean hysteresis length
(corresponding to the burst duration in the Morris-Lecar system)
with increasing noise intensity. The bifurcation diagrams for a
system with and without noise are shown in Fig.~\ref{fig4} (a) and
the distribution of the hysteresis lengths $\Delta r$ for
different noise amplitudes are plotted in Fig.~\ref{fig4} (b). The
 mean hysteresis lengths $\langle \Delta r\rangle $ calculated
 from the distributions reveal the monotonous decrease until they reach
  $\langle \Delta r\rangle =0$. The mean values
 and its standard deviations $\sigma _{\langle \Delta r\rangle}$
 are shown in Fig.~\ref{fig4} (c). This implies the existence of
 a general mechanism of hysteresis reduction induced by noise.

Another example of a system with hysteresis is the equation
describing the dynamics of a pendulum with viscous damping (or a
Josephson junction)~\cite{strogatz}:
\begin{eqnarray}
\dot{\phi }&=&x \nonumber \\
\dot{x}&=&\mu-\sin \phi -\alpha x \label{damp}
\end{eqnarray}
where $\alpha $ and $\mu $ are dimensionless control parameters.
The parameter $\mu$ is assumed to fluctuate as $\mu_t=\mu+D\xi_t$,
where $\mu$ is an average value, $\xi_t$ is a Gaussian white noise
and $D$ is the intensity of the fluctuations. At small $\alpha $
and in the absence of noise ($D=0$), as the parameter $\mu$ is
increased the initially stable fixed point disappears in a
saddle-node bifurcation at $\mu=1$ giving a limit cycle. If $\mu$
is brought back down, the limit cycle persists for $\mu<1$ and its
frequency tends to zero as $\mu=\mu_c$. This change in $\mu$
values as the bifurcation point is reached from different sides
marks the hysteresis region. Now, when the internal parameter
$\alpha $ is made fluctuating, the hysteretic region is seen to
decrease with increasing noise amplitude (see Fig.~\ref{fig4}
(d)). The same phenomenology of hysteresis reduction is observed
also in this case (see Fig.~\ref{fig4} (e)). The averaged
hysteresis depths $\langle \Delta \mu \rangle$ for different noise
amplitudes are shown in Fig.~\ref{fig4} (f).

\subsection{Existence of a minimum mean response duration in the bursting system}

\begin{figure}[h]
\begin{center}
\vspace{3 mm}
(a)\\
\includegraphics*[width=1.\columnwidth]{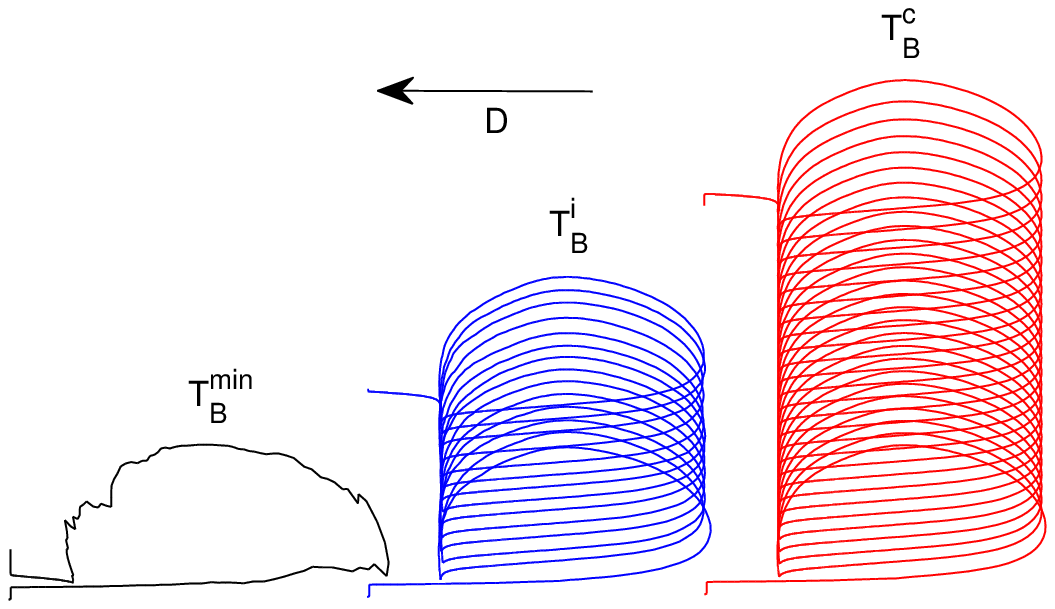}\\
(b)\\
\includegraphics*[width=1.\columnwidth]{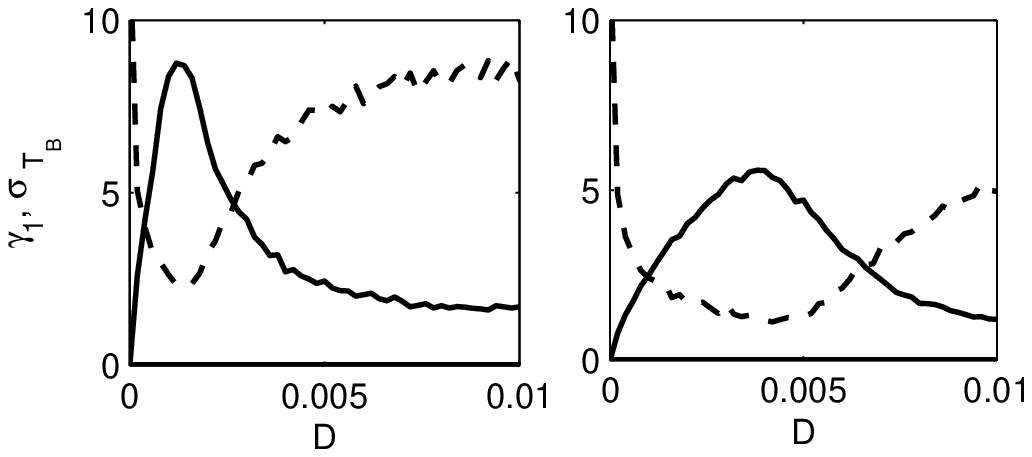}
\caption{\label{fig5} (a) Burst durations in the presence of
noise: minimum possible $T_B^{min}$ (one spike) for $D=0.06$,
intermediate $T_B^{i}$ for $D=0.005$ and in the absence of noise:
$T_B^{c}$ determined by the value of parameter $c$, in this case
$c=0.275$. (b) Correlation between the standard deviation $\sigma
_{T_B}$ (solid line, plotted out of scale) and skewness $\gamma _1
$ (dashed line). Left panel corresponds to Eq.~\ref{eq2} with
$c=0.284$ and the right panel to $c=0.333$.}
\end{center}
\end{figure}

In the case of the Eqs.~(\ref{pitch}) and~(\ref{damp}) the drift
and diffusion processes continue until the averaged value $\langle
\Delta r \rangle =0$ and $\langle \Delta \mu \rangle =0$ are
reached, respectively. Then, when noise still increases, the drift
disappears and only the amplitude of fluctuations around the
states $\langle \Delta r \rangle =0$ and $\langle \Delta \mu
\rangle =0$ increases (i.e. the diffusion increases with
increasing $D$). In the case of a rebound burster, however, in
which the hysteresis is a dynamic process allowing the transient
spiking activity, the burst duration cannot be negative. Thus not
only the drift stops but also the diffusion does. It occurs when
the minimum possible burst duration $T_B^{min}$, being a
single-spike response like in a standard excitable system, is
reached (see Fig.~\ref{fig5} (a)). The maximum variance is
observed when the distribution is situated in the intermediate
distance between two stable states: single spike and a maximum
rebound burst defined at $D=0$. Such a situation permits the
system responses to visit all possible intermediate states giving
the maximal variance. At larger noise amplitudes (but small enough
to keep $FAR=0$), the single spike responses become more probable
giving rise to a $T_B$ variability minimization. The value
$T_B^{min}$ is a kind of a barrier which suppresses further
increase of $\sigma _{T_B}$. The variability increases with $D$
until one tail of a distribution starts to disappear due to a
collision with this barrier. In other words, the maximum
variability of $T_B$ is associated with the symmetry breaking in
the form of distributions for $T_B$'s consisting on the
transitions from a normal to an exponential distribution form (see
Fig.~\ref{fig2} (b)). The symmetry of the probability
distributions can be measured by the skewness coefficient $\gamma
_1$, the standardized moment, defined as follows:
\begin{eqnarray}
\gamma _1 =\frac{\mu _3}{\sigma _{T_B}^3}
\end{eqnarray}
where $\mu _3$ is the third moment about the mean $\langle
T_B\rangle $ and $\sigma _{T_B}$ is the standard deviation. The
skewness of the normal distribution (or any perfectly symmetric
distribution) is zero. The symmetry breaking in the distributions
described by skewness coefficient $\gamma _1$ is shown in
Fig.~\ref{fig5} (b). The correlation between minimum of $\gamma _1
$ (tendency to a normal distribution form) and maximum of $\sigma
_{T_B}$ is clearly seen.

\begin{figure}[h]
\begin{center}
\vspace{3 mm}
\includegraphics*[width=0.8\columnwidth]{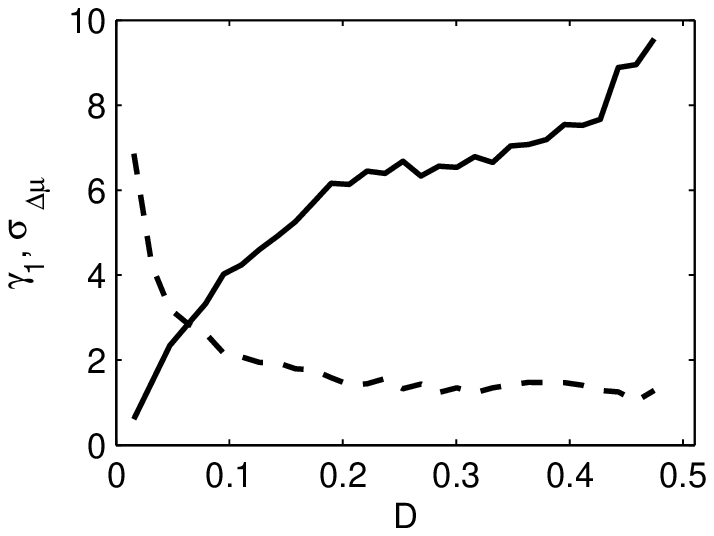}
\caption{\label{fig6} Correlation between the standard deviation
$\sigma _{\Delta \mu}$ (solid line, plotted out of scale) and
skewness $\gamma _1 $ (dashed line) in the case of the equation
describing the dynamics of a pendulum with viscous damping
(Eq.~\ref{damp}).}
\end{center}
\end{figure}

In the case of Eq.~\ref{damp} describing the dynamics of a
pendulum with viscous damping, the distributions for the
hysteresis depth tend to be symmetric as the noise amplitude
increases. This is indicated by the skewness coefficient shown in
Fig.~\ref{fig6}, which at difference to the case of the bursting
system, remains small and constant.

\section{Discussion}

In this work the phenomenon of {\it response} stochastic
incoherence in rebound bursters has been reported. The crucial
ingredients for RSI to appear are the existence of hysteresis in
the system and at least three dimensions, allowing a fast-slow
configuration and rebound bursting regime. The underlying
mechanism has been provided, showing that the maximum variability
in burst durations is caused by the sensitivity of the system
hysteresis responsible for bursting dynamics to noise and by the
symmetry breaking in distributions for the burst durations
occurring as the barrier situated at the minimum value of the
response duration (single spike response) is reached. The
hysteresis vulnerability to noise has been demonstrated on
examples of canonical systems describing hard excitation states.
The implications of this finding is that the bursting responses to
external stimulation are not necessarily well-determined and
fixed, but depend strongly on the slight presence of random
fluctuations. From the physical point of view the phenomenon of
RSI here presented regards a different kind of stochastic
incoherence, namely, the variance of the burst durations instead
of interspike intervals as reported so far in the literature.

The results on the hysteresis reduction could contribute to the
understanding of some phenomena observed in models for bursting
dynamics like the decreasing burst durations in the presence of
noise~\cite{pancreas}. On the other hand, RSI may be relevant in
asynchronous firing patterns as already proposed for the case of
SI in Ref.~\cite{30}.
\section*{Acknowledgments}
The author acknowledges a Marie Curie European Reintegration Grant
(within a 7th European Community Framework Programme).

\end{document}